\documentclass[12pt]{article}
\usepackage{a4wide,latexsym,amsfonts,amsmath,amssymb}
\usepackage[dvips]{graphics}

\newcommand{\eps}{\varepsilon}
\newcommand{\one}{{\bf 1}}
\renewcommand{\sec}[1]{\subsection*{#1}}

\begin{document}

\begin{flushright}  {~} \\[-12mm]
{\sf cond-mat/0404051}\\[1mm]{\sf HU-EP-04/19}\\[1mm]
{\sf Hamburger$\;$Beitr\"age$\;$zur$\;$Mathematik$\;$Nr.$\;$194}
\\[2mm]Phys.\,Rev.\,Lett.\,93 (2004) 070601
\end{flushright} 

\begin{center} \vskip 14mm
{\Large\bf Kramers-Wannier duality from conformal defects}\\[20mm] 
{\large 
J\"urg Fr\"ohlich\,${}^1$,\, J\"urgen Fuchs\,${}^2$,\, 
Ingo Runkel\,${}^3$,\, Christoph Schweigert\,${}^4$}
\\[8mm]
${}^1$ Institut f\"ur Theoretische Physik, ETH Z\"urich, Switzerland\\[1mm]
${}^2$ Institutionen f\"or fysik, Karlstads Universitet, Sweden\\[1mm]
${}^3$ Institut f\"ur Physik, Humboldt Universit\"at zu Berlin, Germany\\[1mm]
${}^4$  Fachbereich Mathematik, Universit\"at Hamburg, Germany
\end{center}
\vskip 22mm

\begin{quote}{\bf Abstract}\\[1mm]
We demonstrate that the fusion algebra of conformal defects of a two-dimensional
conformal field theory contains information about the internal symmetries 
of the theory and allows one to read off generalisations of Kramers-Wannier
duality. We illustrate the general
mechanism in the examples of the Ising model and the three-states Potts model.
\end{quote}
\vfill
\newpage 

\sec{Introduction}

Kramers and Wannier found a high/low temperature duality for the Ising model 
\cite{krwa} that asserts that a correlator of Ising spins 
$\langle \sigma_{x_1} \cdots \sigma_{x_n} \rangle$ at inverse
temperature $\beta$ is equal to a disorder correlation function
$\langle \mu_{x_1} \cdots \mu_{x_n} \rangle$ at the dual inverse temperature 
$\tilde\beta\,{=}\,{-}\,\frac12 \ln \tanh \beta$. In the disorder correlator,
the couplings between neighbouring spins dual to the links of $n/2$ lines,
with each of the positions $x_k$ at the end of one of the lines, are chosen to 
be antiferromagnetic (opposite to the standard ferromagnetic nearest-neighbour 
coupling). This duality has since been considerably generalised, see e.g.\ 
\cite{KWext,DW}.

The significance of Kramers-Wannier duality lies in the fact that it relates
the high-temperature expansion (weak coupling regime) of a lattice model to 
its low-temperature expansion (strong coupling regime) 
and thereby makes the latter accessible to perturbation theory.

Kramers-Wannier-like dualities are also a useful tool in understanding the 
phase structure of a lattice model. At zero magnetic field,
the Ising model has a critical point when $\beta\,{=}\,\tilde\beta$. Its 
universality class is described by a two-di\-men\-sional conformal
field theory (CFT) with central charge $c\,{=}\,\frac 12$.
Physical quantities like critical exponents can then be determined by a CFT 
calculation, relating them to scaling dimensions of bulk fields.
The critical Ising model is self-dual under Kramers-Wannier duality, so 
that a correlator involving spin and disorder
fields is equal to another correlator in the 
same CFT, but with spin fields and disorder fields interchanged.

It is clearly desirable to be able to read off the possible high/low 
temperature dualities leaving a given critical model
fixed solely from knowing its universality class,
i.e., its CFT description. In this letter, we provide such a method by
relating order/disorder dualities of CFT correlators to {\em conformal 
defects\/}. Not every defect can be used to establish a duality, but only 
what we will call `duality defects'. Below we present a method that allows
us to identify such defects by studying the fusion algebra of all conformal 
defects. Duality defects relate perturbations of a CFT in 
different marginal directions, thus allowing one to explore the vicinity
of a model in its moduli space, and they also relate different relevant 
directions, allowing one to extend the order/disorder duality of the CFT to 
a genuine high/low temperature duality away from the critical point.

\sec{Defects in the critical Ising model}

Before exhibiting the underlying mechanism in generality, we investigate
in some detail the critical Ising model as a first non-trivial example.
At central charge 
$c\,{=}\,\tfrac12$ the Virasoro algebra has three unitary irreducible 
highest-weight representations, which we denote by $\one$, $\sigma$, $\eps$. 
Their weights are $h_\one\,{=}\,0$, $h_\sigma\,{=}\,\tfrac1{16}$ and
$h_\eps\,{=}\,\tfrac12$. Correspondingly, there are three primary bulk
fields, the identity $\one$, the spin field $\sigma(z)$
and the energy field $\eps(z)$, with chiral/antichiral conformal weights
$(0,0)$, $(\tfrac1{16},\tfrac1{16})$ and $(\tfrac12,\tfrac12)$, respectively. 

\begin{figure}
\setlength{\unitlength}{1.3pt}
\begin{picture}(300,80)(0,0)
  \put(30,0){
    \put(0,0){\scalebox{1.3}{\includegraphics{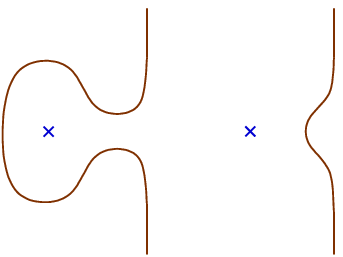}}}
    \put(0,70){ a) } 
    \put(48,35){ $=$ }
    \put(5,27){\scriptsize $T(z)$ }
    \put(64,27){\scriptsize $T(z)$ }
  }
  \put(180,0){
    \put(0,0){\scalebox{1.3}{\includegraphics{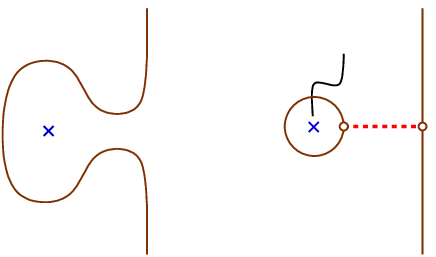}}}
    \put(0,70){ b) }
    \put(44,35){ $\displaystyle = \hspace{-0.5em}
    \underset{{\rm defects}}{\underset{{\rm intermed.}}{\sum}}$ }
    \put(5,27){\scriptsize $\phi(z)$ }
    \put(94,62){\scriptsize $\phi(z)$ }
  }
\end{picture}
\setlength{\unitlength}{1pt}
\caption{A conformal defect is transparent to the stress tensor (a),
while a bulk field $\phi$ generically becomes a sum of
disorder fields (b).}
\label{T-phi-defect}
\end{figure}

Next, we introduce conformal defects. One can think of a conformal defect 
on a surface as being obtained by cutting the surface along the defect line 
and re-joining the two sides of the cut by an appropriate boundary condition, 
i.e.\ a prescription on how bulk fields behave when crossing the cut.
This prescription must preserve the conformal symmetry, i.e.\ both the chiral 
and antichiral components $T(z)$ and $\bar T(\bar z)$ of the conformal stress 
tensor must vary continuously across the cut. In contrast, other bulk fields 
are permitted to exhibit a more complicated behaviour. In fact, dragging a 
conformal defect across a bulk field other than the stress tensor generally 
results in disorder fields, as illustrated in figure \ref{T-phi-defect}.

Because the defect line commutes with the stress tensor, it can be
continuously deformed without changing the value of a correlator.
In this sense a conformal defect is tensionless. Defect lines can only
start and end on field insertions. Such fields
are called {\em disorder fields\/}. Since a defect is invisible
to $T$ and $\bar T$, disorder fields fall into representations of
two copies of the Virasoro algebra, just as the bulk fields do. 

By an argument similar to one used in the analysis of conformal boundary 
conditions \cite{card9}, in the Ising model one finds three conformal defects
\cite{pezu5}. They are labelled by the three $c\,{=}\,\tfrac12$ irreps of 
the Virasoro algebra. The defect of type $\one$ is the trivial defect,
in the presence of which all fields are continuous.
The $\eps$-defect corresponds to a line of antiferromagnetic 
couplings in the lattice
realisation, while the $\sigma$-defect does not have a straightforward
lattice interpretation \cite{Chui} and has long been overlooked.
The appearance of the $\sigma$-defect illustrates that a systematic
analysis of a universality class, using CFT methods,
can lead to structural insight not
obvious from studying a concrete lattice realisation. 

In addition to the bulk fields $\one$, $\sigma(z)$ and $\eps(z)$ we will
also consider the disorder field $\mu(z)$. Pairs of disorder fields 
$\mu(z_1)$ and $\mu(z_2)$ are joined by a defect line of type $\eps$. 
A disorder field has the same conformal weights as the spin 
field, i.e.\ $(\tfrac1{16},\tfrac1{16})$.  

\medskip

\begin{figure}[t]
\setlength{\unitlength}{1.3pt}
\begin{picture}(300,90)(0,0)
  \put(50,0){
    \put(0,0){\scalebox{1.3}{\includegraphics{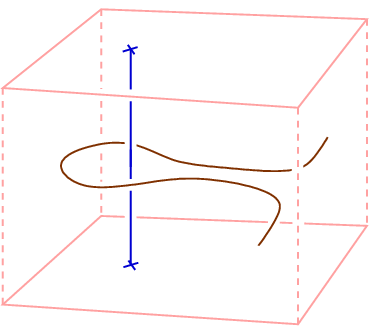}}}
    \put(39,25){\scriptsize $\sigma$ }
    \put(70,48){\scriptsize $\sigma$ }
  }
  \put(170,45){ $=$ }
  \put(200,0){
    \put(0,0){\scalebox{1.3}{\includegraphics{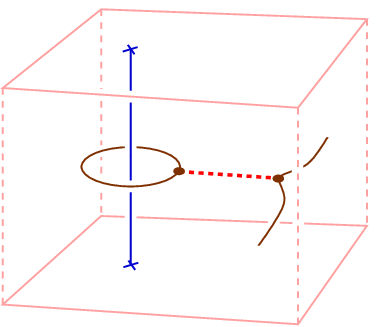}}}
    \put(39,25){\scriptsize $\sigma$ }
    \put(18,43){\scriptsize $\sigma$ }
    \put(92,47){\scriptsize $\sigma$ }
    \put(61,47){\scriptsize $\eps$ }
  }
\end{picture}
\setlength{\unitlength}{1pt}
\caption{The TFT-representation of the pulling a defect
  of type $\sigma$ past a spin field. Collapsing the circular
  $\sigma$-Wilson line on the rhs generates the TFT-representation of the 
  disorder field $\mu(z)$.}
\label{tft-sig-sig}
\end{figure}

The results reported in this letter are obtained in the approach to CFT 
\cite{fffs2,fuRsX} that is based on topological field theory (TFT) in three 
dimensions.  A chiral CFT can be described by the boundary degrees of freedom 
of a three-dimensional topological field theory \cite{witt27,frki}. The 
observables of the TFT are (networks of) Wilson lines. Each 
Wilson line is labelled by a representation of the chiral algebra of the 
CFT, i.e., by $\one$, $\sigma$ or $\eps$ in the example of the Ising model.
The vertices of the network of Wilson lines are labelled
by intertwiners of the corresponding representations. In the TFT formalism
\cite{fffs2,fuRsX}, a CFT correlator on a surface $X$ (oriented, without 
boundary) with field insertions is described as follows: one first constructs 
a three-manifold by taking an interval above each point of $X$, 
$M \,{=}\, X \,{\times}\, [1,-1]$. 
The two boundary components $X \,{\times}\,\{1\}$ and $X \,{\times}\, \{-1\}$ 
support the two chiral degrees of freedom of the CFT, respectively. 
At each field insertion on $X$, a Wilson line with the corresponding label
is inserted which runs along the interval $[-1,1]$, thus connecting the two 
boundary components of $M$. 
A defect line on the surface $X$ is described by a Wilson line 
inserted on $X \,{\times}\, \{0\} \,{\subset}\, M$ and labelled again by 
$\sigma$ or $\eps$, depending on the defect type. 
Consider, for instance, the effect of pulling a $\sigma$-defect past a spin 
field $\sigma(z)$ as in figure \ref{T-phi-defect}b. This turns out to 
generate a disorder field $\mu(z)$ and an $\eps$-defect. 
In the TFT formalism, this process amounts to the
identity in figure \ref{tft-sig-sig}, which is then easily verified.

\medskip

\begin{figure}[b]
\setlength{\unitlength}{1.2pt}
\begin{picture}(300,120)(0,5)
  \put(0,70){
    \put(13,0){
      \put(0,0){\scalebox{1.2}{\includegraphics{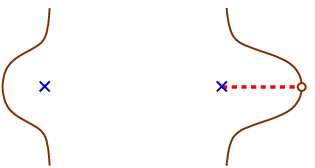}}}
      \put(15,38){\scriptsize $\sigma$ }
      \put(72,36){\scriptsize $\sigma$ }
      \put(75,26){\scriptsize $\eps$ }
      \put(9,16){\scriptsize $\sigma(z)$ }
      \put(52,14){\scriptsize $\mu(z)$ }
    }
    \put(0,40){ a) }
    \put(47,23){ $=$ }
    }
  \put(140,70){
    \put(-2,0){
      \put(0,0){\scalebox{1.2}{\includegraphics{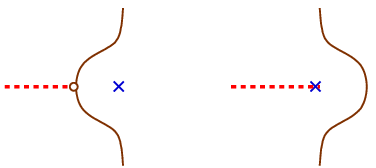}}}
      \put(5,18){\scriptsize $\eps$ }
      \put(74,18){\scriptsize $\eps$ }
      \put(37,39){\scriptsize $\sigma$ }
      \put(95,39){\scriptsize $\sigma$ }
      \put(32,17){\scriptsize $\sigma(z)$ }
      \put(84,16){\scriptsize $\mu(z)$ }
    }
    \put(0,40){ b) }
    \put(47,23){ $=$ }
    }
  \put(280,70){
    \put(17,0){
      \put(0,0){\scalebox{1.2}{\includegraphics{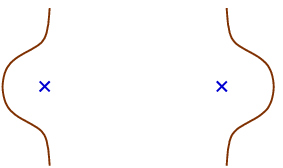}}}
      \put(15,38){\scriptsize $\sigma$ }
      \put(72,36){\scriptsize $\sigma$ }
      \put(14,17){\scriptsize $\eps(z)$ }
      \put(46,15){\scriptsize $-\eps(z)$ }
    }
    \put(0,40){ c) }
    \put(47,23){ $=$ }
    }
  \put(0,0){
    \put(15,0){
      \put(0,0){\scalebox{1.2}{\includegraphics{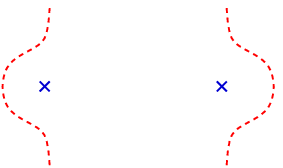}}}
      \put(15,38){\scriptsize $\eps$ }
      \put(72,36){\scriptsize $\eps$ }
      \put(14,17){\scriptsize $\sigma(z)$ }
      \put(46,15){\scriptsize $-\sigma(z)$ }
    }
    \put(0,40){ d) }
    \put(47,23){ $=$ }
    }
  \put(140,0){
    \put(13,0){
      \put(0,0){\scalebox{1.2}{\includegraphics{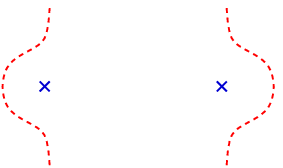}}}
      \put(15,38){\scriptsize $\eps$ }
      \put(72,36){\scriptsize $\eps$ }
      \put(14,17){\scriptsize $\eps(z)$ }
      \put(48,17){\scriptsize $\eps(z)$ }
    }
    \put(0,40){ e) }
    \put(47,23){ $=$ }
    }
  \put(280,0){
    \put(10,0){
      \put(0,0){\scalebox{1.2}{\includegraphics{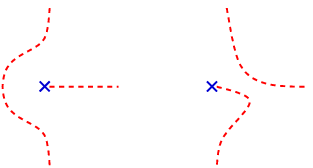}}}
      \put(15,39){\scriptsize $\eps$ }
      \put(68,39){\scriptsize $\eps$ }
      \put(65,4){\scriptsize $\eps$ }
      \put(25,25){\scriptsize $\eps$ }
      \put(9,17){\scriptsize $\mu(z)$ }
      \put(51,16){\scriptsize $\mu(z)$ }
    }
    \put(0,40){ f) }
    \put(47,23){ $=$ }
    }
\end{picture}
\setlength{\unitlength}{1pt}
\caption{Taking defects of type $\sigma$ and $\eps$
past field insertions. The TFT-representation of a) is given in 
figure \ref{tft-sig-sig}.}
\label{ising-rules}
\end{figure}

A straightforward calculation within the TFT-frame\-work allows one to find the set 
of rules summarised in figure \ref{ising-rules} for taking defects past field
insertions. In this figure, the normalisation of the fields is chosen such that
$\langle \sigma(z)\, \sigma(w) \rangle \,{=}\,
\langle \mu(z)\, \mu(w) \rangle \,{=}\, |z{-}w|^{-1/4}$
and $\langle \eps(z) \,\eps(w) \rangle \,{=}\,
   $\linebreak[0]$
   |z{-}w|^{-2}$.
Also, three-valent vertices between two $\sigma$-de\-fects and one 
$\eps$-defect have been labelled with a suitably normalised intertwiner.

\begin{figure}[bt]
\begin{picture}(300,120)(0,0)
  \put(0,130){ a) }
  \put(0,5){
    \put(0,0){\includegraphics{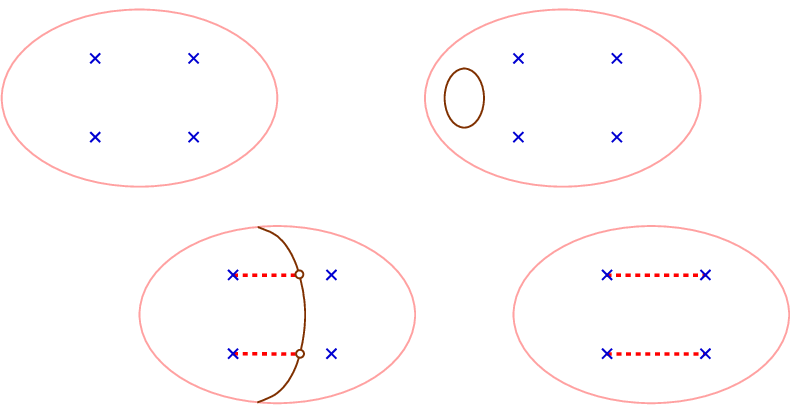}}
    \put(82,88){ $\displaystyle =  \frac{1}{\sqrt{2}}$ }
    \put(2,24){ $\displaystyle = \frac{1}{\sqrt{2}}$ }
    \put(126,24){ $\displaystyle = $ }
    \put(0,0){
      \put(25,104){\scriptsize $\sigma$ }
      \put(25, 71){\scriptsize $\sigma$ }
      \put(54,104){\scriptsize $\sigma$ }
      \put(54, 71){\scriptsize $\sigma$ }
    }
    \put(122,0){
      \put(25,104){\scriptsize $\sigma$ }
      \put(25, 71){\scriptsize $\sigma$ }
      \put(54,104){\scriptsize $\sigma$ }
      \put(54, 71){\scriptsize $\sigma$ }
    }
    \put(40,-63){
      \put(25,104){\scriptsize $\mu$ }
      \put(25, 71){\scriptsize $\mu$ }
      \put(54,104){\scriptsize $\sigma$ }
      \put(54, 71){\scriptsize $\sigma$ }
    }
    \put(147,-63){
      \put(25,104){\scriptsize $\mu$ }
      \put(25, 71){\scriptsize $\mu$ }
      \put(54,104){\scriptsize $\mu$ }
      \put(54, 71){\scriptsize $\mu$ }
    }
    \put(141,87){\scriptsize $\sigma$ }
    \put( 89,24){\scriptsize $\sigma$ }
    \put( 75,33){\scriptsize $\eps$ }
    \put( 75,17){\scriptsize $\eps$ }
    \put(187,33){\scriptsize $\eps$ }
    \put(187,17){\scriptsize $\eps$ }
  }
  \put(240,130){ b) }
  \put(240,0){
    \put(0,0){\includegraphics{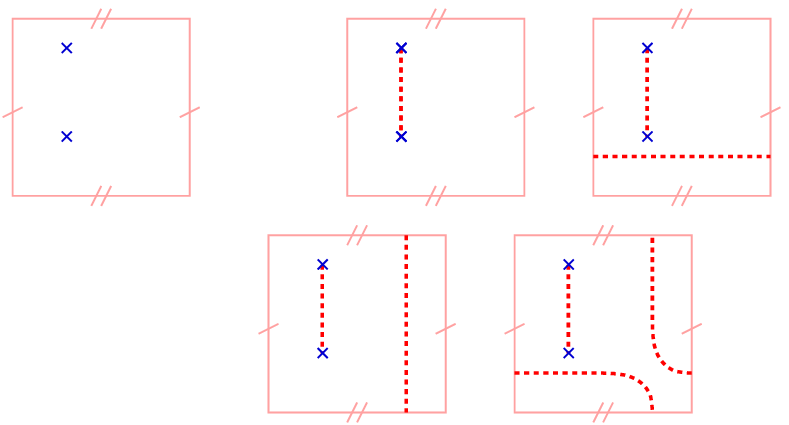}}
    \put(58,88){ $\displaystyle = \, \frac{1}{2} \Bigg($ }
    \put(153,88){ $\displaystyle +$ }
    \put(58,24){ $\displaystyle +$ }
    \put(130,24){ $\displaystyle +$ }
    \put(208,24){ $\displaystyle \Bigg)$ }
    \put(207,72){\scriptsize $\eps$ }
    \put(119,42){\scriptsize $\eps$ }
    \put(190,42){\scriptsize $\eps$ }
    \put(154, 9){\scriptsize $\eps$ }
    \put(0,0){
      \put(96,45){\scriptsize $\mu$ }
      \put(96,20){\scriptsize $\mu$ }
      \put(87,32){\scriptsize $\eps$ }
    }
    \put(72,0){
      \put(96,45){\scriptsize $\mu$ }
      \put(96,20){\scriptsize $\mu$ }
      \put(87,32){\scriptsize $\eps$ }
    }
    \put(-74,62){
      \put(95.5,43){\scriptsize $\sigma$ }
      \put(95.5,17.5){\scriptsize $\sigma$ }
    }
    \put(23,62){
      \put(96,45){\scriptsize $\mu$ }
      \put(96,20){\scriptsize $\mu$ }
      \put(87,32){\scriptsize $\eps$ }
    }
    \put(94,62){
      \put(96,45){\scriptsize $\mu$ }
      \put(96,20){\scriptsize $\mu$ }
      \put(87,32){\scriptsize $\eps$ }
    }
  }
\end{picture}  
\caption{Order/Disorder duality of a correlator of four spin
fields on a sphere, and of two spin fields on a torus,
as induced by the $\sigma$-defect.}
\label{ising-ex} \end{figure}

We can now obtain the first example for an order/disorder duality, 
the correlator of four spin fields on the sphere. In this correlator
we insert a small circular $\sigma$-defect, which changes the value
of the correlator by the quantum dimension 
${\rm dim}(\sigma) \,{=}\, \sqrt{2}$ of the representation $\sigma$.
Pulling the defect circle around the sphere and past the field insertions
results in a disorder correlator as shown in figure \ref{ising-ex}a.
Using the rules of figure \ref{ising-rules}, it is easy to verify that
repeating this procedure removes the $\eps$-defects and replaces
the disorder fields again by spin fields. This is the order/disorder
duality of the Ising model on the sphere. 
The rules in figure \ref{ising-rules} immediately imply that 
when one studies duality on a torus, the non-trivial topology 
results in a sum over several configurations, as illustrated in 
figure \ref{ising-ex}b.

The mechanism can be generalised to surfaces with boundaries. In the Ising 
model, the boundary conditions are again
labelled by the $c\,{=}\,\tfrac12$ irreps \cite{card9}: $\one$ and $\epsilon$ 
describe fixed boundary conditions with `spin up' and `spin down', respectively,
while $\sigma$ describes the `free' boundary condition. Owing to the Ising 
fusion rules $\sigma \,{\star}\, \one\,{=}\, \sigma\,{\star}\, \epsilon
\,{=}\, \sigma$ and $\sigma \,{\star}\, \sigma\,{=}\,\one\,{+}\,\eps$,
a $\sigma$-defect in front of a `spin up' or a `spin down' boundary condition
can be replaced by a `free' boundary condition without defect,
while a $\sigma$-defect in front of a `free' boundary condition yields the 
sum of a `spin up' and a `spin down' boundary condition.  One
thus obtains the well-known duality of fixed and free boundary conditions
\cite{DW}.

\medskip

So far, we have considered the order/disorder duality only at the critical 
point. However, the rules listed in figure \ref{ising-rules} also allow 
us to establish the duality away from the critical point.
For example, note that taking a $\sigma$-defect through the
energy field $\eps(z)$ results in a change of sign. Perturbing the CFT by 
$\eps(z)$ amounts to a change of temperature, and applying the duality to 
each term in a perturbation series leads to the equality
$$ 
  \big\langle 
  \, \sigma(x)\, \sigma(x') \, 
  {\rm e}^{-\lambda \int\! \eps(y) \,{\rm d}^2y} 
  \,\big\rangle 
  = 
  \big\langle 
  \, \mu(x)\, \mu(x') \, 
  {\rm e}^{\lambda \int\! \eps(y) \,{\rm d}^2y} 
  \,\big\rangle 
$$
for the example of a two-point correlator on the sphere.

\sec{The general mechanism}

We are now in a position to describe a general mechanism that works for all
unitary rational conformal field theories. For such models, there
is a finite set of primary bulk fields $\phi_a(z)$.
One denotes the number of such fields transforming in representations $i$ and 
$j$ of the chiral and antichiral symmetries, respectively, by $Z_{ij}$. The 
matrix $Z$ thus describes the modular invariant torus partition function of 
the CFT.

We restrict our attention to conformal defects that preserve enough additional 
symmetry to keep the model rational. We call a defect `simple', iff it cannot 
be written as a sum of other defects. The number of simple defects
is given by ${\rm tr}(Z Z^{\rm t})$ \cite{pezu5,fuRsX}. Let us denote the 
set of simple defects by $\{ D_\alpha \,|\, \alpha\,{\in}\,\mathcal{K} \}$ 
for some label set $\mathcal{K}$, 
with the label for the trivial defect denoted by `$e$'.
In general, one must assign an orientation to a defect line. 

Consider two simple defects running parallel
to each other and with the same orientation. In the limit of vanishing
distance they fuse to a single defect which is, in general,
a superposition of simple defects. 
This gives rise to a (not necessarily commutative) fusion algebra of
defects \cite{pezu5,oc-fusion}, written schematically as
$$
  D_\alpha \otimes D_\beta 
  = \sum_{\gamma\in\mathcal{K}} \hat N_{\alpha\beta}^{~~\gamma}\,
  D_\gamma \,.
$$
In the TFT formalism, the general class of models we are studying now 
is described by an algebra $A$ in the category of representations of the 
chiral algebra of the CFT. Defects are then described as bimodules of $A$,
and the defect fusion rules above amount to decomposing
the tensor product over $A$ of two bimodules into a direct sum of simple
bimodules, which can be performed explicitly. The bimodule describing
the trivial defect $D_e$ turns out to be $A$ itself.
If the two parallel defects have opposite direction we write
$D_\alpha \,{\otimes}\, D_\beta^\vee$.

Two subsets of defects turn out to be of particular interest. The
first one is the set $\mathcal{G}$ of {\em group-like\/} defects. A
defect $X$ is called group-like, iff $X{\otimes}X^\vee \,{=}\, D_e$.
One can show that group-like defects are simple, so 
that $\mathcal{G} \,{\subseteq}\, \mathcal{K}$. Further,
for two group-like defects $D$ and $D'$, their fusion $D\,{\otimes}\,D'$ is 
again group-like. This turns $\mathcal{G}$ into a (in general nonabelian) 
group with unit $D_e$, via
$D_g\,{\otimes}\,D_h \,{=}\, D_{gh}$ and $D_{g^{-1}} \,{=}\, D_g^\vee$.
{}From figure \ref{T-phi-defect}b we see that taking
any group-like defect past a bulk field results in
a sum of bulk fields, since the only intermediate defect
that does occur is the trivial one,
$D_g\,{\otimes}\, D_g^\vee \,{=}\, D_e$. Commuting a group-like
defect past all bulk fields in a correlator results in 
a correlator of different bulk fields, but having the same value. Thus,
{\em group-like defects produce an internal symmetry of the CFT\,}.
For the Ising model one has $\mathcal{G} \,{=}\, \{\one,\eps \}$, a
$\mathbb{Z}_2$ group, and from figure \ref{ising-rules}d we see that 
the defect $\eps$ indeed acts by reversing the sign of the spin field.

The second and larger subset is formed by the
{\em duality defects\/}. A defect $X$ is a duality defect,
iff there exists another defect $Y$ such that taking first 
$X$ and then $Y$ past a bulk field results only in a sum
of bulk fields, with no disorder fields present.
In other words, commuting $X$ past all fields in an order correlator
in general gives a disorder correlator. However, subsequently 
commuting $Y$ past all fields in this disorder correlator gives 
back an order correlator.
Thus, {\em duality defects produce order-disorder dualities of the CFT}.
Using the TFT formalism, one can establish the following simple 
characterisation of duality defects: $X$ is a duality defect, if and only 
if every simple defect in $X \,{\otimes}\, X^\vee$ 
is a group-like defect. A detailed proof will be presented elsewhere.
Clearly, the set $\mathcal{D}$ of simple duality defects satisfies
$\mathcal{G} \,{\subseteq}\, \mathcal{D} \,{\subseteq}\, \mathcal{K}$.

Note that in order to determine $\mathcal{G}$ and $\mathcal{D}$ in a given
model, it suffices to know the fusion algebra of defects.
In the Ising model one finds
$\mathcal{D} \,{=}\, \{ \one, \sigma, \eps \} \,{=}\, \mathcal{K}$.
The duality defect $\sigma$ generates the original Kramers-Wannier duality.

The above discussion is limited to the critical point.
However, suppose that for a given duality defect $D_\alpha$ we can 
find a bulk field $\phi(z)$ such that taking $D_\alpha$ past $\phi(z)$ results
in another bulk field $\tilde\phi(z)$, rather than in a sum of bulk fields and 
disorder fields. (In the Ising model, the field $\eps(z)$ has this property 
with respect to the defect labelled by $\sigma$, see figure \ref{ising-rules}c.)
Then the duality induced by $D_\alpha$ provides an equality between a correlator
of the CFT perturbed by $\int\! \phi(z)\, {\rm d}^2z$ and the dual
correlator perturbed by $\int\! \tilde\phi(z)\, {\rm d}^2z$.

\sec{The critical three-states Potts model}

The critical three-states Potts model has central charge $c\,{=}\,4/5$ and 
corresponds to a $D$-type model in the classification 
of Virasoro-minimal models. It has first been considered in \cite{dots}.
The number of simple conformal defects in this model is
${\rm tr}(Z Z^{\rm t})\,{=}\, 16$ 
(and there are 8 conformal boundary conditions).
The defect fusion rules can be computed using Ocneanu quantum algebras
\cite{pezu5,oc-fusion}, or weak Hopf algebras, or by TFT methods. The result 
can be summarised as follows. The set of defect labels can be written as 
$\mathcal{K} \,{=}\, \mathcal{K}_x \,{\times}\, \mathcal{K}_y$
with $\mathcal{K}_x \,{=}\, S_3 \,{\cup}\, \{ u_+, u_- \}$ and
$\mathcal{K}_y \,{=}\, \{\one,\varphi\}$, where 
$S_3$ denotes the permutation group of three symbols.
The fusion product $D_{x,y} \,{\otimes}\, D_{x',y'} \,{=}\, 
\sum_{r \in x \cdot x'} \sum_{s \in y \cdot y'}\! D_{r,s}$ is
obtained by the following rules. The product in $\mathcal{K}_y$
is given by Lee-Yang fusion rules 
$\varphi \,{\cdot}\, \varphi \,{=}\, \one + \varphi$, while the product
in $\mathcal{K}_x$ is described as follows. For $p,p' \,{\in}\, S_3$,
$p \,{\cdot}\, p'$ is given by the product in $S_3$, and
$p \,{\cdot}\, u_\eps \,{=}\, u_{\eps'}$ with $\eps\,{\in}\,\{\pm 1\}$ and 
$\eps'\,{=}\, \eps\, {\rm sgn}(p)$; finally, denoting the elements of $S_3$ 
by $e$ (identity), $p_{12}$, $p_{13}$, $p_{23}$ (transpositions), and
$p_{123}$, $p_{132}$ (cyclic permutations), we have 
$u_+\,{\cdot}\,u_+ \,{=}\, u_-\,{\cdot}\,u_- \,{=}\, e + p_{123} + p_{312}$
and $u_+\,{\cdot}\,u_-\,{=}\,u_-\,{\cdot}\,u_+ \,{=}\, p_{12}+p_{13}+p_{23}$.
Owing to the presence of $S_3$, the fusion algebra of defects is 
non-commutative in this model.  One can convince oneself that the group-like 
defects are $\mathcal{G} \,{=}\, \{ (p,\one) \,|\, p \,{\in}\, S_3 \}$ and 
the duality-defects are $\mathcal{D} \,{=}\, \{ (x,\one) \,|\, x \,{\in}\,
\mathcal{K}_x \}$.

The $S_3$-structure of the group-like defects could again have
been expected from the lattice model realisation of the three-states
Potts model; it amounts to a permutation of the three possible values
of the spin.

The critical three-states Potts model contains 12 primary 
bulk fields and 208 primary disorder fields. Of these,
we consider the energy operator $E(z)$ of left/right conformal weight
$(\frac25,\frac25)$, the two spin fields $S_\pm(z)$ of
weight $(\frac1{15},\frac1{15})$ and the two disorder fields
$Z_{\pm}(z)$ of the same weight, where $Z_+$ generates a
defect of type $(p_{123},\one)$ and $Z_-$ one of type $( p_{132},\one )$. 
We find that taking a duality defect of type $(u_\eps,\one)$ through a 
spin field $S_\nu(z)$, for $\eps,\nu\,{\in}\,\{\pm 1\}$, 
generates a disorder field $Z_{\eps\nu}(z)$,
and vice versa. Furthermore, taking $D_{u_\pm,\one}$ past
the energy field $E(z)$ gives $-E(z)$, so that the the order/disorder
duality at the critical point extends to a high/low temperature
duality off the critical point.

\sec{Conclusions}

We have demonstrated that the fusion algebra of defects in a CFT contains 
a lot of physical information: Internal symmetries correspond to group-like 
defects, and the order/disorder dualities to duality defects. 
The analysis is carried out within CFT, it allows one to study symmetry 
properties of universality classes of critical behaviour without reference 
to a particular lattice realisation. 
To compute the dual correlator one must simply commute a
given duality defect past all field insertions. This procedure
can be applied to correlators on surfaces of arbitrary genus and even with 
boundary. Via conformal perturbation theory, one can also identify high/low 
temperature dualities in the vicinity of the critical point.

To conclude, we mention that these considerations can also be
applied to the free boson. One then finds that $T$-duality
is induced by duality defects, too. The defect line in this example is
labelled by the $\mathbb{Z}_2$-twisted representation 
of the $U(1)$-current algebra.

\vskip3em

 \def\anop  {Ann.\wb Phys.}
 \def\comp  {Comm.\wb Math.\wb Phys.}
 \def\ijmp  {Int.\wb J.\wb Mod.\wb Phys.\ A}
 \def\jopa  {J.\wb Phys.\ A}
 \def\jgap  {J.\wb Geom.\wB and\wB Phys.}
 \def\nupb  {Nucl.\wb Phys.\ B}
 \def\phlb  {Phys.\wb Lett.\ B}
 \def\phrl  {Phys.\wb Rev.\wb Lett.}
 \def\phrv  {Phys.\wb Rev.}
 \def\remp  {Rev.\wb Mod.\wb Phys.}
 \newcommand\wb{\,\linebreak[0]} \def\wB {$\,$\wb}
 \newcommand\Bi[1]    {\bibitem{#1}}
 \newcommand\J[5]     {{#1} {#2} ({#3}) {#4}}
 \newcommand\K[6]     {{#1} {#2} ({#3}) {#4}}

\end{document}